\documentclass[fleqn,usenatbib]{mnras}

\usepackage{newtxtext,newtxmath}

\usepackage[T1]{fontenc}

\DeclareRobustCommand{\VAN}[3]{#2}
\let\VANthebibliography\thebibliography
\def\thebibliography{\DeclareRobustCommand{\VAN}[3]{##3}\VANthebibliography}

\usepackage{graphicx}	
\usepackage{amsmath}	

\title[The variable stars in the cluster NGC 5634]{The variable stars in the compact halo globular cluster NGC 5634 visited again}

\author[Arellano Ferro et al.]{
A. Arellano Ferro$^{1}$\thanks{E-mail: armando@astro.unam.mx}, R. Michel$^{2}$, M.A. Yepez$^{3}$, S. Muneer$^{4}$, I. Bustos Fierro$^{5}$, Z. Prudil$^{6}$
\\
$^{1}$Instituto de Astronom\'ia, Universidad Nacional Aut\'onoma de M\'exico, Ciudad de M\'exico, CP 04510, M\'exico.\\
$^{2}${Observatorio Astron\'omico, Nacional, Instituto de Astronom\'ia, Universidad Nacional Aut\'onoma de M\'exico, Ensenada, Mexico.}\\
$^{3}${Instituto Nacional de Astrof\'isica, \'Optica y Electr\'onica (INAOE), Luis Enrique Erro No.1, Tonantzintla, Pue., C.P. 72840, M\'exico}\\
$^{4}${Indian Institute of Astrophysics, Bangalore, India.}\\
$^{5}${Observatorio Astron\'omico, Universidad Nacional de C\'ordoba, C\'ordoba C.P. 5000, Argentina.}\\
$^{6}${European Southern Observatory, Karl-Schwarzschild-Straße 2, 85748, Garching, Germany}\\
}

\date{Accepted -- 27. Received --; in original form --}

\pubyear{2020}

\begin{document}
\label{firstpage}
\pagerange{\pageref{firstpage}--\pageref{lastpage}}
\maketitle


\begin{abstract}
We present new time-series CCD \emph{VR} photometry of the globular cluster NGC 5634. We aim to use the known RR Lyrae stars, members of the cluster, as indicators of mean metallicity and distance. Accurate coordinates, periods and an identification chart of the variables in the field of our images are provided. A membership analysis was performed, based on $Gaia$-DR3 proper motions, for 3525 point sources within 15 arcmin from the cluster center. The membership status for each known variable was established and it was found that V10, V11 and V16 are most likely field stars. The variability of the RRab star V7, considered non-variable for a number of years, is demonstrated. Via the Fourier decomposition of the light curves of cluster member RR Lyrae, the mean metallicity and distance were calculated independently from RRab and RRc stars to find [Fe/H]$_{\rm ZW}= -1.67 \pm 0.11$ y $D=23.9 \pm 0.8$ kpc,
and [Fe/H]$_{\rm ZW}= -1.69 \pm 0.22$ y $D=22.9 \pm 1.0$ kpc respectively.
\end{abstract}



\section{Introduction}

The globular cluster NGC 5634  (C1427-057 in the IAU nomenclature) ($\alpha = 14^{\mbox{\scriptsize h}}
29^{\mbox{\scriptsize m}} 37.23^{\mbox{\scriptsize s}}$, $\delta = -05^{\circ}58\arcmin
35.1\arcsec$, J2000; $l = 342.21^{\circ}$, $b =+49.24^{\circ}$) 
is a high galactic latitude outer halo object at some 25 kpc away from the Sun and about 21 kpc from the Galactic center \citep{Harris1996}. It has been scarcely studied in the past. It is a difficult target since there is a very bright foreground star at only $\sim$ 1.3 arcmin from the cluster center, obliging to some maneuvers to avoid it at the time of both the observations and reductions.

\begin{figure*}
\begin{center}
\includegraphics[width=17.5cm]{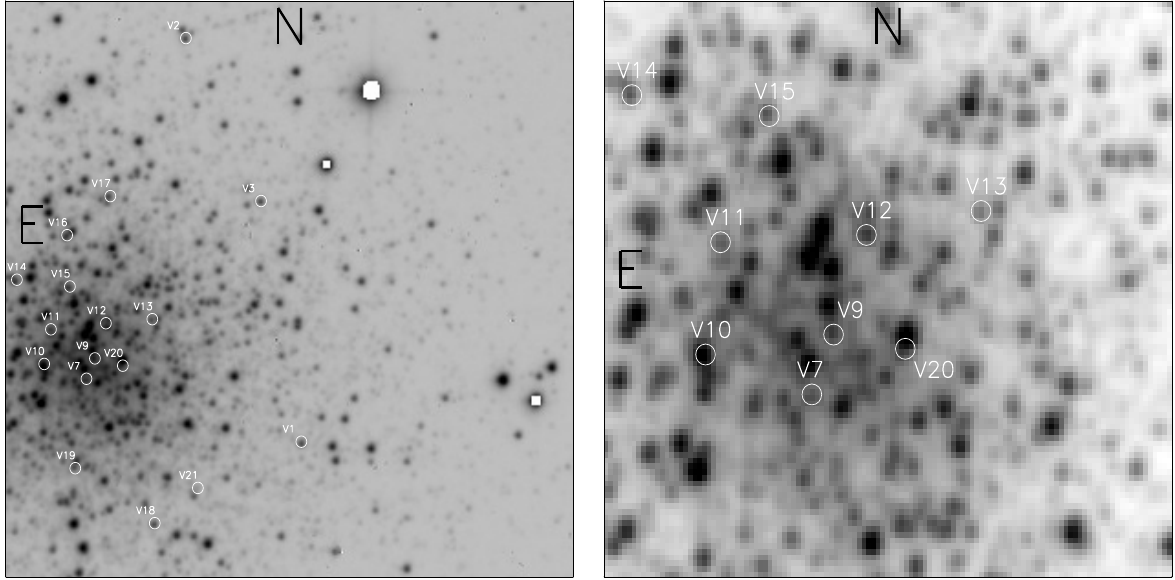}
\caption{Identification chart of the variable stars in the field of NGC 5634. The left panel is approximately 2.5$\times$2.5 arcmin square. To facilitate the identification of the central region the panel to the right is  about 0.45$\times$0.45 arcmin square.}
\label{ID_CHART}
\end{center}
\end{figure*}

Based on its position and radial velocities, \citet{Bellazzini2002} have associated the cluster with the merging of the dwarf spheroidal galaxy Sagittarius with the Milky Way, suggesting that NGC 5634 was an original resident of the trapped galaxy.  The cluster appears as a highly concentrated system. Its core and half-light radii (see \citet{Harris1996}), which at a distance of $\sim$ 25 kpc, are $R_c \sim$ 0.6 pc and $R_h \sim$ 6 pc, respectively, support the idea that the cluster lost its external halo due to tidal stripping forces \citep{Salinas2005}.

The population of variable stars in the field of NGC 5634 was discovered by \citet{Baade1945} (V1-V7) and 60 years later by \citet{Salinas2005} (V8-V21).
The positions and periods of these variables are listed in the Catalog of variable stars of globular clusters (CVSGC) \citep{Clement2001}. 

There are reasons to pay a new visit to the small population of variable stars in this cluster; the coordinates listed in the CVSGC for some variables were based on relative (X,Y) positions measured on photographic plates, and an approximation of the cluster center. Being such a compact and faint cluster and having nearly half of its variables buried deep in the central regions, it is clear that some variables are heavily blended, hence a reconsideration of their coordinates and variable type may be useful. 
At present we have access to new tools that facilitate the identification of the exact position of a variation, for instance one can perform a blinking of the full collection of astrometrically calibrated differential images to spot where the variations take place. 
Given the advent of accurate proper motions and parallaxes from $Gaia$-EDR3 \citep{Lindegren2021}, it has been shown by \citet{Prudil2024}, based on the membership analysis performed for 170 globular clusters \citep{Vasiliev2021}, that some of those variables are in fact field stars (see also the membership flag in \citet{Clement2001}). 
In the present paper we shall report the results and caveats of our time series photometry, and we will perform a new stellar membership by an alternative approach. We shall produce and discuss a $V$ vs $V-R$ colour-magnitude diagram (CMD) and will use for the first time the recovered light curves of member variables towards independent estimations of the metallicity and distance from their Fourier decomposition.

\section{Observations and image reductions}
\label{sec:ObserRed}

\subsection{Observations and reductions}

All observations were performed with two telescopes; the 1.5m telescope in San Pedro M\'artir Observatory, Baja California, Mexico, during 5 nights between May 1 and May 11, 2024. The detector used was a 2048x2048 pixel
CCD (e2v CCD-231-42) (Marconi 5) with an approximate scale of 0.493 arcsec per pixel after a binning of 2x2 pixels. A subarray of 700 x 700 pixels was used, for a field of approximately 3.45 x 3.45 arcmin square. Tyical exposure times were 120-600 seconds in $V$ and 80-120 seconds in $R$. The second telescope was the Himalayan Chandra 2.0m telescope of the Indian Astrophysical Observatory (IAO) at Hanle, India on March 3 and April 4, 2024. The telescope was equipped with an E2V CCD44-82-0-E93 detector (2048×4096) with a plate scale of 0.296 arcsec per pixel. A subarray of 1140 x 800 pixels was employed, corresponding to a field of view of about  5.6×3.9 arcmin square. Only a handful of $V$ images were obtained with this instrument. Typical exposure times were 600 seconds in $V$.

For image photometric treatment, we employed the  Difference Image Analysis (DIA) technique to extract photometry of all point sources in our images of the cluster. We used the \emph{DanDIA} pipeline \citep{Bramich2008,Bramich2013,Bramich2015}, which has been described in detail by \citet{Bramich2011}.
 The \emph{DanDIA} pipeline creates a reference image for each filter by stacking a set of best-seeing images taken on a single night. A sequence of difference images was created by subtracting the reference image from each image in the time series collection. Differential fluxes are measured for each point source in the field, and instrumental light curves are built. The complete process has been described in detail in Section 2 of \citet{Arellano2008}.

The presence of the $\sim$ 7 mag star HD 127119 at about 1.3 arcmin of the cluster center hinders the accurate photometry in the whole field of the cluster. To avoid contamination of our images by "light bleeding", we decided to shift the cluster in the field of our images and leave the bright star off the chip. The price we had paid for this maneuver was not observing the variables V4, V6, and V8. The variable V5 was inadvertently lost at the southern edge of our field of observation. The remaining known variables are within our observed field and are identified in the ID chart in Fig. \ref{ID_CHART}. The majority of the known variable stars are well within the half-light radius of the cluster. 

\begin{figure}
\begin{center}
\includegraphics[width=7.9cm]{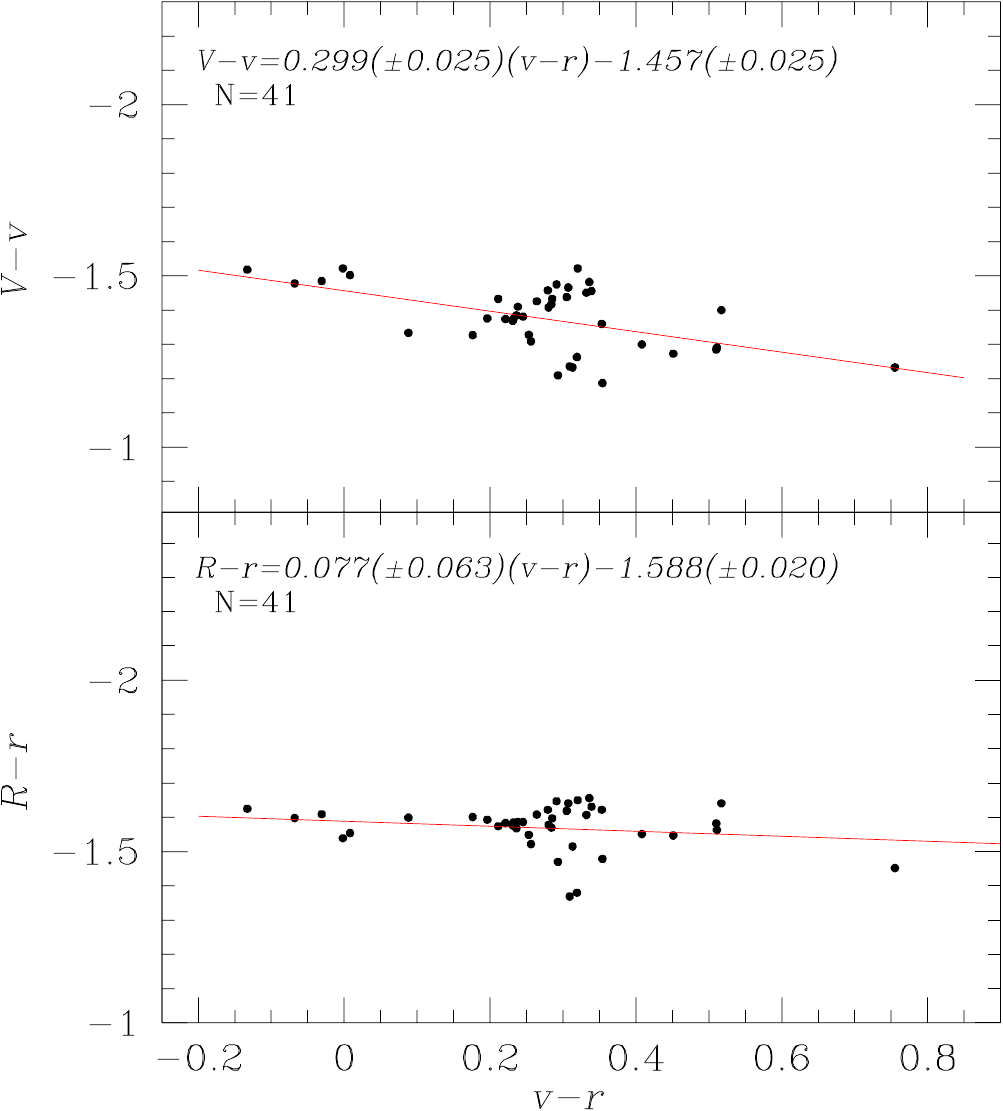}
\caption{Transformation relationship between instrumental and standard $VR$ photometric systems. The transformations were defined employing 42 local standard stars from the collection of \citet{Stetson2000}.}
\label{Trans_color}
\end{center}
\end{figure}

\subsection{Transformation to the Standard System}
\label{Tranformation}

The instrumental photometry was transformed  to the standard  
Johnson-Kron-Cousins photometric system \citep{Landolt1992}
\emph{VR}, using local standard stars in the fields of the target clusters. These standard stars have been taken from the extensive collection of
\citet{Stetson2000}\footnote{%
 \texttt{https://www.canfar.net/storage/list/STETSON/Standards}}. We found 41 standard stars with instrumental light curves in the field of our images. The transformation of the instrumental light curves into the standard system was done independently for the SPM and Hanle data. Being our SPM data more abundant and taken under better seeing conditions, we a posteriority applied small offsets to the Hanle $V$ data to match the SPM light curves. The colour $(v-r)$ dependence of the standard minus instrumental magnitudes, $V-v$ and $R-r$, are shown in Fig. \ref{Trans_color} for the SPM data. The corresponding transformation equations are also given in the figure.

\section{Stellar membership analysis}
\label{sec:membership}

We performed a stellar membership analysis in a grand field of NGC 5634, using for such an aim the high-quality proper motions available in the third data release of the $Gaia$ mission \citep{Gaia2023}.

\begin{figure*}
\begin{center}
\includegraphics[width=16.0cm]{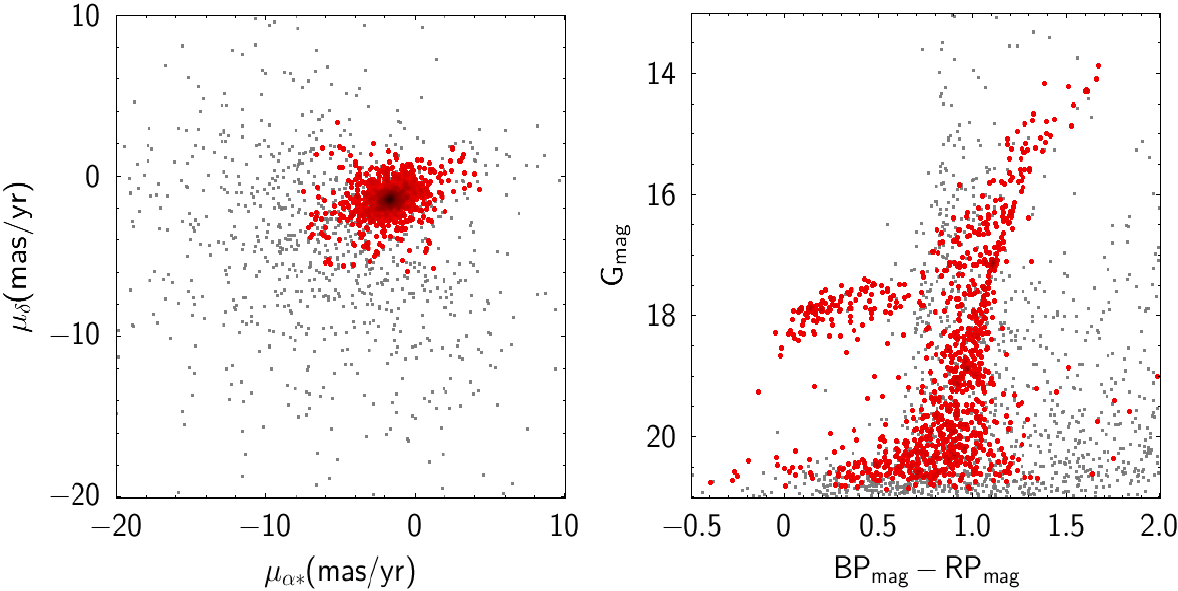}
\caption{$Gaia$-DR3 VPD (left panel) and CMD (right panel) of the cluster NGC 5634. Red and gray points correspond to likely cluster members and field stars respectively, determined as described in section \ref {sec:membership}. A total of 3525 $Gaia$ point sources within 15 arc minutes are displayed, while 1357, were found to be cluster members, all of them contained with a radius of 4 arcmin. }
\label{VPD}
\end{center}
\end{figure*}

We used the method developed by \citet{Bustos2019} which is based on a two-step approach: 1) to find groups of stars with similar characteristics in the four-dimensional space of the gnomonic coordinates ($X_{\rm t}$,$Y_{\rm t}$) and proper motions ($\mu_{\alpha*}$,$\mu_\delta$) employing the BIRCH clustering algorithm \citep{Zhang1996} and 2) in order to extract likely members that were missed in the first stage, the analysis of the projected distribution of stars with different proper motions around the mean proper motion of the cluster is performed.

The analysis was carried out on 3525 $Gaia$ point sources within 15 arc minutes from the center of NGC 5634 and 1357 were found to be likely cluster members. The Vector Point Diagram (VPD) and the Colour Magnitude Diagram (CMD) are shown in Fig. \ref{VPD}, where field stars (gray dots) and cluster member stars (red dots) are distinguished. It is interesting to note that the farthest member is at about 4 arcmin from the cluster center, which at a distance of about 23 kpc corresponds to a distance projected on the plane of the sky of about 27 pc confirming the compact nature of this halo cluster.

\begin{figure*}
\begin{center}
\includegraphics[width=16.0cm]{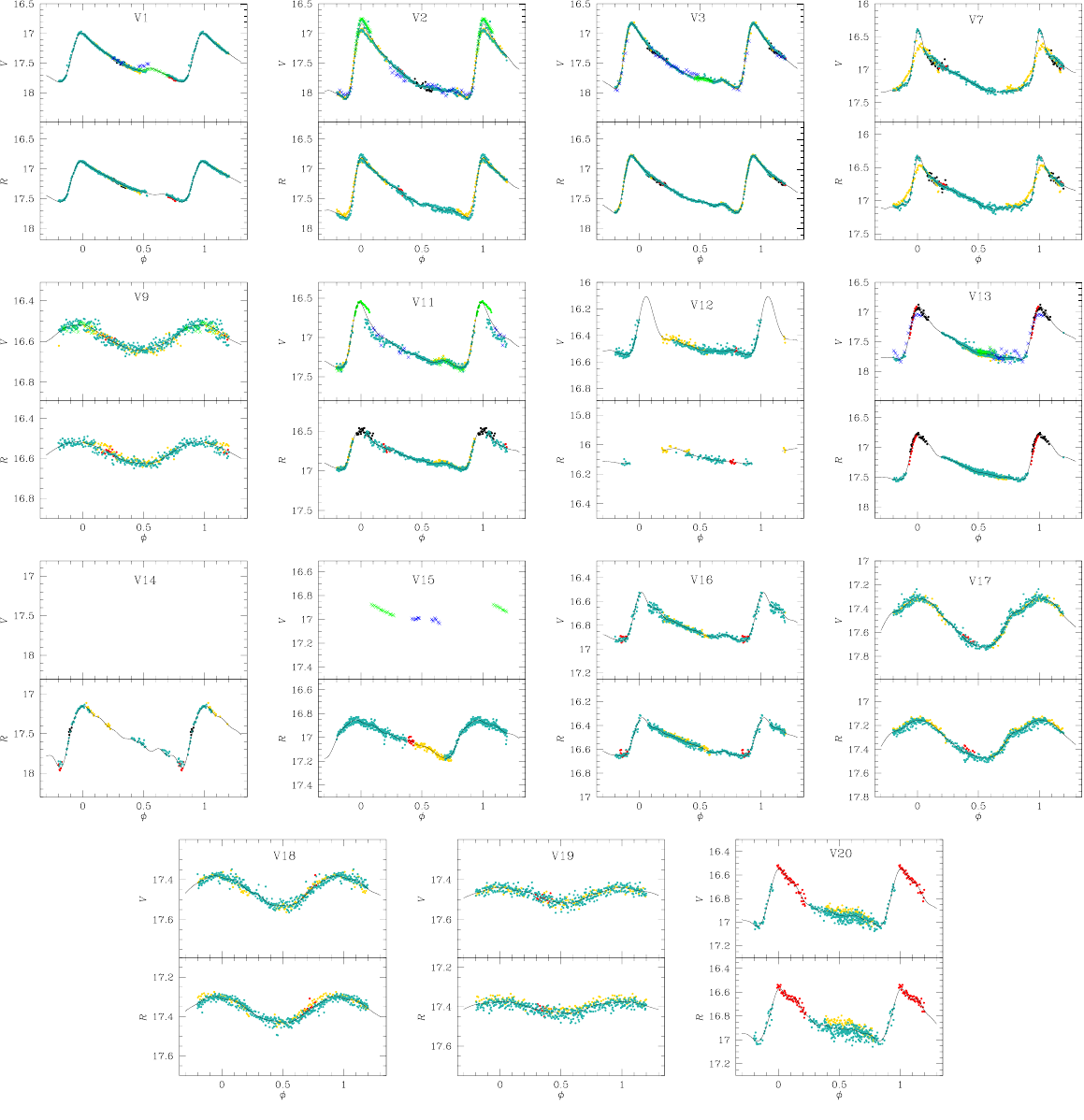}
\caption{$V$ and $R$ light curves of variable stars in the field of NGC 5634. Colour code is as follows: red, yellow and dark green dots are SPM data obtained between May 01 and 11 2024. Green and blue crosses are $V$ data obtained on March 14 and April 4 respectively, at the IAO.}
\label{LightCurves}
\end{center}
\end{figure*}

\begin{figure}
\begin{center}
\includegraphics[width=8.0cm]{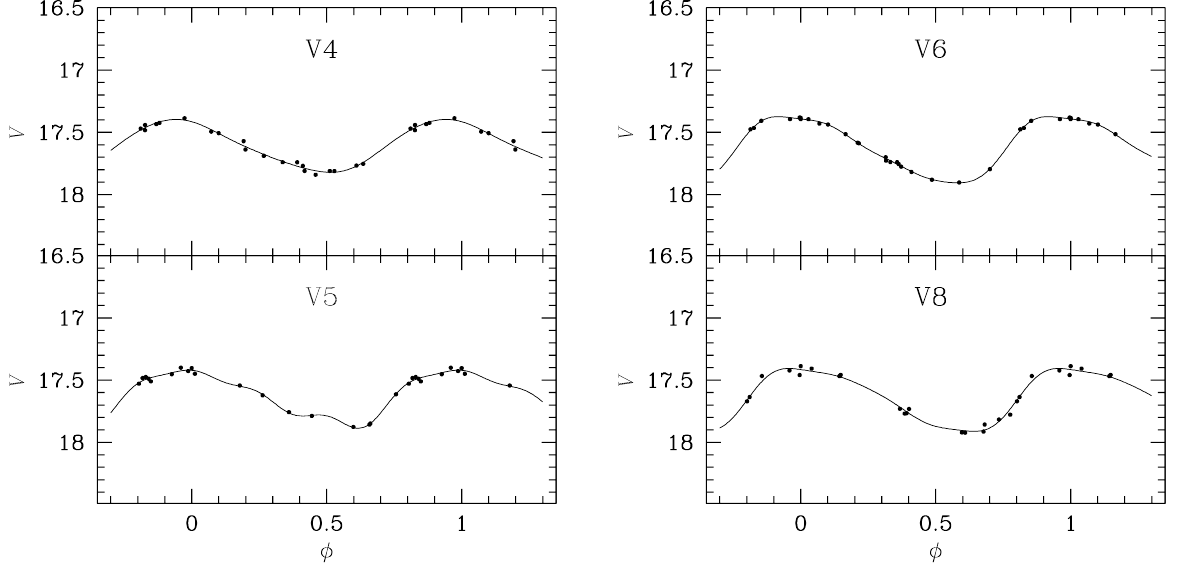}
\caption{$Gaia$ $G$-band light curves, converted into $V$-band \citep{Riello2021},  of 4 RRc stars}
\label{4GAIA}
\end{center}
\end{figure}

Before evaluating the cluster membership of the known variables, we revised their coordinates. We proceeded by blinking all our best quality differential images in small regions around each variable to confirm the temporal variation and the precise position of the blinking source. We noted some small differences with the coordinates published in the present edition of the CVSGC. In Table 1 we list all intensity weighted mean magnitudes, amplitudes and periods for variables star measured in this study. We also include their coordinates and membership status.

From this membership analysis, we found that among the known variables V11, and V16 are likely field stars. The membership flag for variables in the CVSGC,  that emerges from an independent membership analysis, \citep[see][]{Prudil2024}, also indicates the field nature of V11 and V16.
Nonetheless, the two analyses disagree with respect to the nature of V20, but considering that this star is heavily blended, its membership status is indeed dubious. We shall return to the membership status of V20 in Sections 4.2  and 7.

\begin{table*}

\small
\label{tab:datosgenerales1}
\centering
    \caption{General data of the variables in the FoV of NGC 5634.}
    \begin{tabular}{cccccccccccc}
    \hline
    Variable &$Gaia$ &Type & $<$\textit{V}$>$ & $<$\textit{R}$>$ & $A_{V}$ & $A_{R}$ & Period & $HJD_{max}$ & RA & Dec. & Membership \\
     & id  & &(mag) & (mag) & (mag) & (mag) & (days) & (d+2460000) & (J2000.0) & (J2000.0) & (m/f)\\
    \hline
    V1&{\it3641359074110990464} & RRab & 17.483 & 17.265  & 0.826 & 0.697 &0.657848  & 0441.8987 & 14:29:33.51 & -05:58:56.72 & m \\
    V2 &{\it3641359275974954496}  &RRab & 17.657 &17.430   & 1.331 & 1.085 & 0.605147 & 0441.9460 & 14:29:35.67 & -05:57:14.83 & m \\
    V3 &{\it3641359177190278528} & RRab & 17.534 & 17.367  & 1.106 & 0.976 & 0.600419 & 0441.9433 & 14:29:34.30 & -05:57:55.52 & m \\

    V4 &{\it3641358902313485824}& RRc &17.617  &  --   &0.481  & --   & 0.385674 &  --  & 14:29:40.88&  -05:59:43.98 & m \\

    V5 &{\it3641355427683683456}& RRc &17.635  &  --   & 0.481 & -- & 0.406517 & --   & 14:29:36.38&  -06:01:20.72& m \\

    V6 &{\it 3641358902312004352}& RRc &17.637  & --    &0.519  & --   & 0.344183 & --  & 14:29:40.16 & -05:59:31.19 & m \\

    V7 &3641358971033097344& RRab & 17.094 & 16.906 &0.968  &0.828  & 0.581274 & 0436.9661 & 14:29:37.23 & -05:58:42.02 & m \\ 

V8 &{\it3641358902312006400} & RRc &17.653  & --    &0.547  & -- & 0.330169 & --   & 14:29:41.11&  -05:59:16.63 & m \\    
         
    V9 &3641358971032981888& RRc & 16.585  & 16.523 & 0.132 & 0.110 & 0.317521 & 0441.9460 & 14:29:37.12 & -05:58:36.54 & m \\
    V10 &3641358971032992768 & RRab & 15.1:   & 14.6:  & -- & -- & -- & -- & 14:29:37.99 & -05:58:38.69 & m \\
    V11 &3641358971033107072 & RRab & 17.118 &16.788 & 0.827 & 0.506 & 0.663947 & 0403.8276 & 14:29:37.87 & -05:58:29.39 & f \\
    V12 &3641358971031515904 & RRab & 16.451 & 16.082 & 0.40 & 0.12: & 0.624556 & 0441.9928 & 14:29:36.90 & -05:58:28.38 & m \\
    V13 &{\it3641359177191448064} & RRab & 17.509 &17.280  & 0.949 & 0.778 & 0.645648 & 0413.8012 & 14:29:36.15 & -05:58:26.27 & m \\
    V14 &3641358971031788672 & RRab & --  & 17.543 & -- & 0.785 & 0.725933 & 0434.8970 & 14:29:38.51 & -05:58:17.38 & m \\
    
    V15 &3641358971033084672 & RRab & 16.89:  & 17.000 & -- & 0.338 & 0.859346 & 0441.8137& 14:29:37.58 & -05:58:18.75 & m \\
    
    V16 &3641359211549981312  & RRab & 16.797 &16.538 & 0.368 & 0.335 & 0.669811 & 0441.9873 & 14:29:37.66 & -05:58:05.66 & f \\
    
    V17 &{\it3641359211550005632} & RRc & 17.509 &17.317  & 0.421 & 0.334 & 0.405502 & 0441.8219 & 14:29:36.93 & -05:57:55.30& m \\
    
    V18 &{\it3641358867952297600} & RRc &17.456 &17.366 & 0.159 & 0.147 & 0.325762& 0432.9253 & 14:29:36.02& -05:59:18.17& m \\
    
    V19 &3641358971031744640 & RRc & 17.477 &17.403  & 0.082 & 0.073 & 0.295919 & 0441.8110 & 14:29:37.41&-05:59:04.12 & m \\
    
    V20 &3641358971033060864 & RRab & 16.857 &16.839  & 0.514 & 0.463 & 0.645748 & 0431.8422 & 14:29:36.64& -05:58:37.94 & m *\\

    V21 &3641358867952534656 & SX Phe & --   &--  &--  & --&0.0666 & 0431.8422 & 14:29:35.29& -05:59:08.6 & m\\    
    \hline
     G1 &{\it 3641359177191556096} &   &  &  &  & & &  & 14:29:35.10&  -05:57:58.97 & m\\
     
     G2 &{\it3641358867952547712} &   &  &  &  & & &  & 14:29:35.00 & -05:58:58.33& m\\
     
     G3 &{\it3641358971033104384} &   &  &  &  & & &  & 14:29:38.17 & -05:58:49.99 & m\\
     
     G4 &{\it3641355947375985792} &   &  &  &  & & &  & 14:29:43.16 & -05:59:38.64& m\\
     G5 &{\it3641358971033112960} &   &  &  &  & & &  & 14:29:39.92 & -05:58:39.44 & m\\
     G6 &{\it3641358971033056384} &    &  &  &  & & &  & 14:29:36.88 & -05:58:26.47 & m\\
    \hline
    \end{tabular}

    \center{Stars with $Gaia$ id numbers in italics, are varaibles reported in $Gaia$-DR3.\\

     Columns 3 and 4 are intensity weighted mean magnitudes, columns 5 and 6 are light curves amplitudes. Column 11 indicates the cluster membership status; m - members, f - field.\\

    * The membership status of V20 is discussed in section 4.2}

\end{table*}

\section{Variable star light curves recovery and caveats}
\label{Vars}

Most known variable stars in NGC 5634 are highly concentrated within the half-light radius of the cluster, and bad blending with multiple neighbours of similar magnitude or brighter is common. This anomalous circumstance, together with the image scale of our instruments and the poor seeing conditions that prevailed during some of our observing nights, prevented us from isolating and properly measuring some variables. In Fig. \ref{LightCurves} we display the $VR$ light curves of those stars we were able to measure during our best quality nights, which are colour-coded as described in the caption. In spite of the above limitations the displayed light curves are of quality enough to perform a Fourier decomposition analysis aimed at estimating the distance and metallicity, as shall be discussed later in this work.

All of our \emph{VI} photometry for the studied variables in this work is provided in Table \ref{tab:vr_phot}. A small portion of this table is given in the printed version of this paper and the full table is available in electronic form in the {\it Centre de Donnes astronomique de Strasbourg} data base (CDS). 

\begin{table}
\footnotesize
\caption{Time-series $V$ and $R$ photometry for all the variables in the field of view of NGC 5634. (Full table is available in electronic format).}
\centering
\begin{tabular}{ccccc}
\hline
Variable &Filter & HJD & $M_{\mbox{\scriptsize std}}$ &
$\sigma_{m}$  \\
Star ID  &        & (d) & (mag) &(mag)\\
\hline
V1& V & 2460403.79371&17.153& 0.009\\
V1& V & 2460403.79517&17.186& 0.008\\
\vdots   & \vdots & \vdots  & \vdots & \vdots  \\
V1& R & 2460413.76279&17.375& 0.007\\
V1& R & 2460413.76545&17.378& 0.007 \\
\vdots   & \vdots & \vdots  & \vdots & \vdots  \\
V2& V & 2460403.79371&17.328& 0.010\\
V2& V & 2460403.79517&17.272& 0.009\\
\vdots   & \vdots & \vdots  & \vdots & \vdots \\
V2& R & 2460431.83550&17.325& 0.007\\
V2& R & 2460431.83823&17.323& 0.007\\
\vdots   & \vdots & \vdots  & \vdots & \vdots \\
\hline
\end{tabular}
\label{tab:vr_phot}
\end{table}

At this point, a few comments on the variables V7 and V10 are in order. 

V7 was discovered by \citet{Baade1945} together with the other six variables V1-V6. V7 is the most central of the seven stars and while properly identified it was not measured. Later the star was not detected as variable by \citet{Salinas2005} and hence was listed as constant in the CVSGC. We have been able to isolate and measure the star on the best sample of our images. The variability was confrmed using the same technique as Baade; the blinking-comparison, but in this case of the many differential images built by DanDIA. The variation is conspicuous. The light curve is included in Fig. \ref{LightCurves}. The star is clearly a RRab star with amplitude modulations of the Blazhko type and has a period of 0.581274 d.

V10, on the other hand, was discovered by  \citep{Salinas2005} where the diplayed light curve is clearly that of a RRab star. Using the blinking technique, we confirmed the variability at a position very similar to the one reported in the CVSGC. Unfortunately, DanDIA was unable to extract a proper light curve due to the extreme blending conditions at that point. However, the star residing at that position is very bright, the mean $VR$ magnitudes place the star near the tipo of the RGB in the CMD. Our conclusion is that if indeed we are dealing with an RRab star, then it is not a cluster member, but rather sits in front of the system.

\subsection{Variables in the $Gaia$-DR3}

The $Gaia$-DR3 reports 16 variables in the field of NGC 5634 and are included in Table \ref{tab:datosgenerales1} with their $Gaia$ identification in italics. Ten of them match the known variables in the CVSGC. We found $G, G_{BP}$ and $G_{RP}$ light curves for six stars \citep{Clementini2023}; these correspond to the known RRab variables V2 and V3, and to the four RRc variables V4, V5, V6 and V8.
Stars with $Gaia$ id number in bold font in Table \ref{tab:datosgenerales1} were missed by the $Gaia$ mission because these stars are in the central region of the cluster, they are blended with brighter stars or are too faint.

As mentioned in Section \ref{sec:ObserRed}, the RRc stars V4, V5, V6 and V8 are not in the field of our images. In order to include them in the Fourier decomposition analysis, we converted the $G$-band magnitudes into $V$-band, using the correlations of \citet{Riello2021}. These light curves are shown in Fig. \ref{4GAIA}. Their physical parameters will be presented in the following section.

The remaining six $Gaia$ variables are listed in the bottom part of the table labeled "G". Stars G4 and G5 are outside the field of view of our images. G3 was not recovered by our photometry since it is badly blended and too close to the edge of our images. G1, G2, and G6, are declared as eclipsing binaries by $Gaia$ and we do have photometry for the three of them. However, even employing our best-quality data, i.e. the same sets used for the RR Lyrae in Fig. \ref{LightCurves}, we do not detect any variation in either of the three. Thus, we could not confirm their variation. 

\subsection{A brief comment on the Oosterhoff type of NGC 5634}
\label{par:Bailey}

\begin{figure}
\begin{center}
\includegraphics[width=8.3cm]{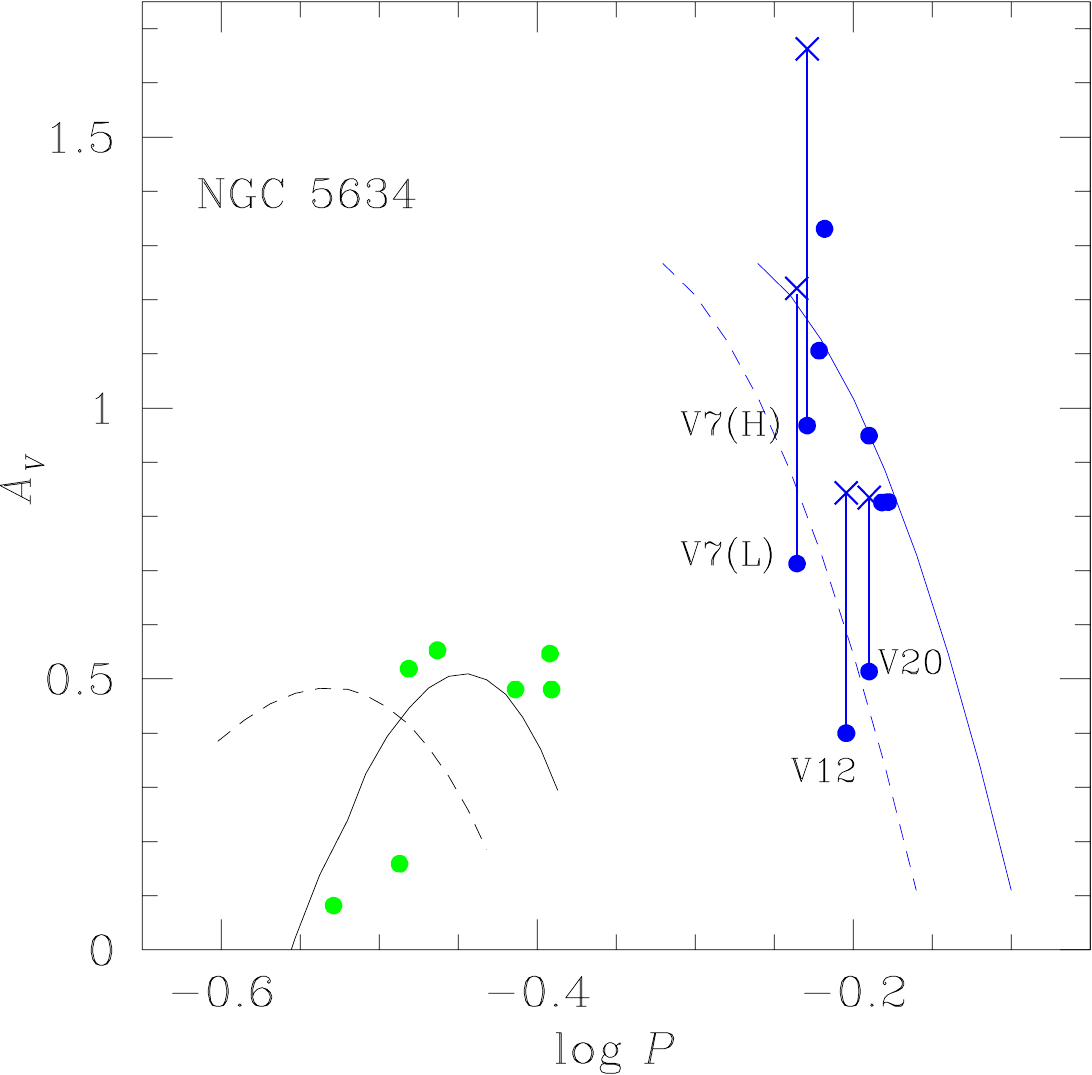}
\caption{The period-amplitude or Bailey's diagram for NGC 5634. Blue and green symbols are used for RRab and RRc stars respectively. Blue dashed and solid loci represent the distribution of unevolved and evolved RRab star typical of Oo I and Oo II clusters respectively according to \citet{Cacciari2005}. Black loci are the equivalent distributions for RRc according to \citet{Arellano2015} and \citet{Kunder2013a}. The star distribution clearly point NGC 5634 as an Oo II type cluster. The amplitudes of V7, V12 and V20 were corrected (blue crosses) for the presence of unresolved neighbours. See Section \ref{par:Bailey} for a discussion. }
\label{BAILEY}
\end{center}
\end{figure}

\begin{figure*}
\begin{center}
\includegraphics[width=12.0cm]{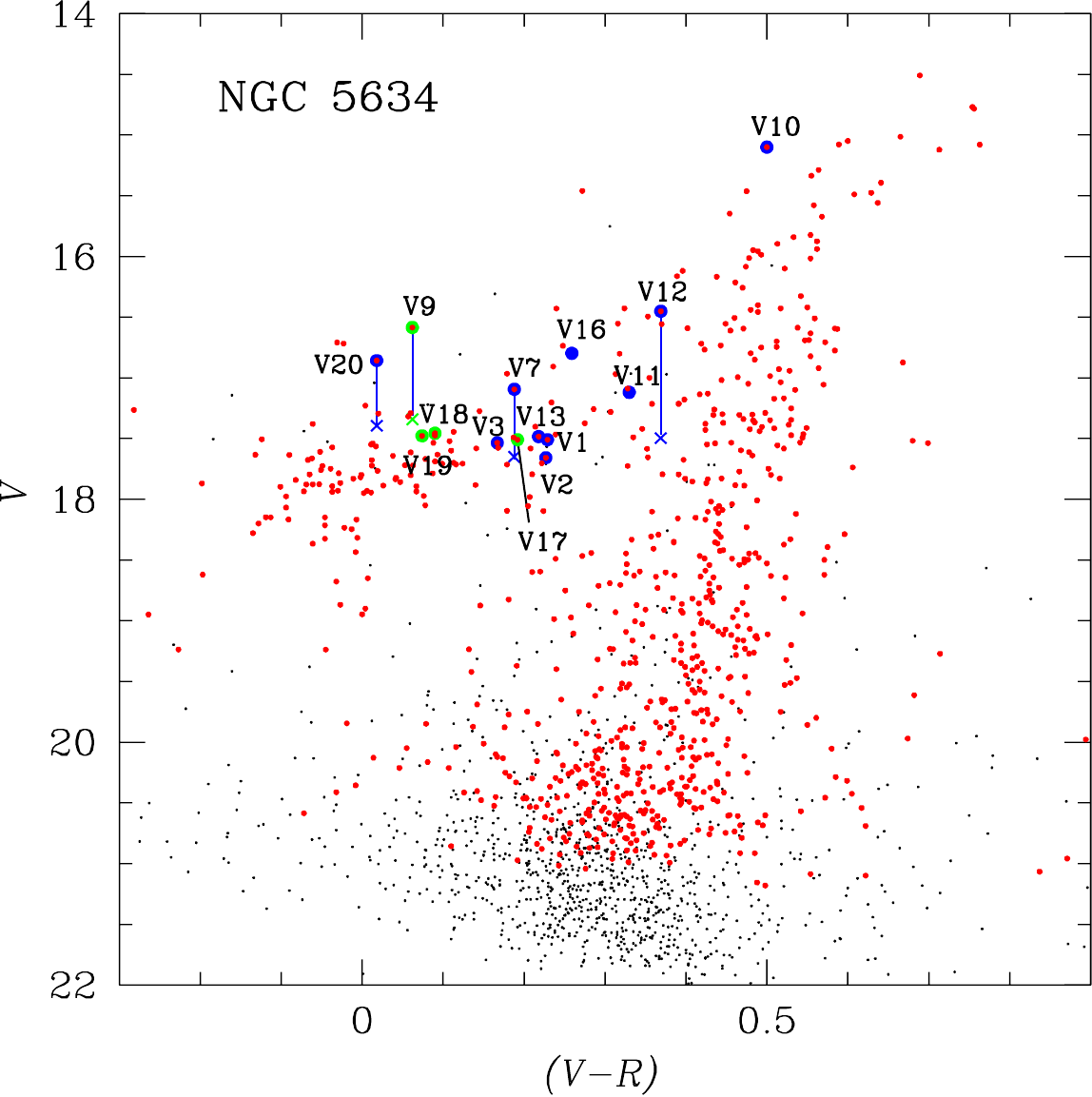}
\caption{The NGC 5634 $V -(V-R)$ Colour-Magnitude diagram. Red points are stars found as likely cluster members while gray dots represent the field star. Blue and green circles are used for RRab and RRc stars respectively. Hence, cluster member variables have a red dot in the center. Positions of the member stars V9, V12, V15 and V20, originally off the HB, could be corrected from the presence of a well identified close neighbour. The odd position of the aparently cluster member V10 could not be deblended. The Fourier physical parameters listed in Table \ref{tab:parfisRR} were calculated only for member stars in the HB.}
\label{CMD}
\end{center}
\end{figure*}

With the estimated period and amplitudes for the RRab and RRc stars given in Table \ref{tab:datosgenerales1}, the resulting log $P$ - amplitude diagram, or Bailey's diagram, is displayed in Fig. \ref{BAILEY} where the solid loci show the distribution of RRab and RRc stars in Oo II clusters, in opposition to the dashed loci typical of Oo I type clusters. The star distributions of NGC 5634 clearly identify the cluster as an Oo II. Three outstanding stars in the diagram are V7, V12 and V20 whose amplitudes seem smaller than expected. We found that very close to the position of these variables, there are two $Gaia$ stars with comparable brightness separated less than 1", therefore unresolved in the ground-based observations. In the cases of V7 and V12 we have determined via a careful blinking of the residual images, that the actual variables are the brighter to the west and the fainter to the east of each pair respectively. For the V20 pair, the variable is clearly the one to the southeast, whose light curve is that of a RRab with $V$ $\sim 16.8$ mag (see Fig. \ref{LightCurves}). We calculated that V7 appears 0.580 mag brighter, V12 appears 0.559 mag brighter and V20 0.524 mag brighter due to the blending with the unresolved neighbours. These corrections would place V7, V12 and V20 closer to the HB in the CMD (Fig. \ref{CMD}).

With the same assumptions, we also calculated that the actual amplitude of V12 should be 0.503 mag instead of 0.300 as measured (blended).  However, the light curve of V12 is missing the maximum. We estimated its amplitude from a Fourier fit but it is possible that the amplitude has been underestimated. For V20 the amplitude should be about 0.835 mag instead of 0.514 (blended). Nevertheless, for V7, it should be noted that its light curve displays amplitude modulations (see Fig. \ref{LightCurves}), hence we corrected both the low and high extreme amplitudes for the presence of the unresolved companion, V7(L) and V7(H) in Fig. \ref{BAILEY}. These corrections would place them closer to the solid blue curve (Oo II cluster) in the Bailey's diagram.

The average periods for the RRab and RRc stars in Table \ref{tab:datosgenerales1} are: $<P_{ab}> = 0.66 \pm 0.08$ and $<P_{c}> = 0.35 \pm 0.074$ which also indicate that NGC 5634 is of the Oo II type. Another indicator of the Oosterhoff type in globular clusters is the frequency of RRc stars relative to the total number of RR Lyrae, i.e. NRRc/(NRRab+NRRc), for which, from the data in Table \ref{tab:datosgenerales1}, we get 0.5. This clearly classifies the cluster as OoII \citep[see Fig 5.]{Castellani2003}.

\section{Fourier decomposition of RR Lyrae light curves}
\label{sec:Four}

The Fourier RR Lyrae light curve decomposition towards the estimation of stellar physical parameter is a well established approach. The technique and the relevant semi-empirical calibrations for the calculation of metallicity, luminosity, radius and mass have been described in detail in the papers by \citet{Arellano2022,Arellano2024}. 

The light curve is fitted with a number of harmonics of the Fourier series model of the form:

\begin{equation}
\label{eq.Foufit}
m(t) = A_0 + \sum_{k=1}^{N}{A_k \cos\ ({2\pi \over P}~k~(t-E) + \phi_k) },
\end{equation}

\noindent
where $m(t)$ is the magnitude at time $t$, $P$ is the period, and $E$ is the epoch. A
linear
minimization routine is used to derive the 
best-fit values of the 
amplitudes $A_k$ and phases $\phi_k$ of the sinusoidal components. 
From the amplitudes and phases of the harmonics in eq.~\ref{eq.Foufit}, the 
Fourier parameters, defined as $\phi_{ij} = j\phi_{i} - i\phi_{j}$, and $R_{ij} =
A_{i}/A_{j}$, are computed. The Fourier coefficients are not specifically listed in this paper but they are available on request.

The low-order Fourier parameters can be used in combination with
semi-empirical calibrations to calculate stellar values of [Fe/H], $M_V$, mass, and radius. The physical parameters for the measured RR Lyrae stars in NGC 5634 are listed in Table \ref{tab:parfisRR}.

\begin{table*}
\footnotesize
\centering 
\caption[]{Physical parameters from the member RR Lyrae Fourier light curve decomposition.}
    \begin{tabular}{ccccccccc} 
    \hline
    ID &[Fe/H]$_{\rm ZW}$ & [Fe/H]$_{\rm UVES}$  & $M_V$ & log~$T_{\rm eff}$  &log$(L/{L_{\odot}})$ &$M/{ M_{\odot}}$ & $D(kpc)$ &$R/{ R_{\odot}}$ \\
    \hline
    \multicolumn{9}{c}{RRab}\\
    \hline
    V1  &-1.50(3) & -1.41(3)   & 0.51(1) & 3.80(1) & 1.71(1) & 0.66(7) & 23.09(3) & 5.94(1) \\ 
    V2 &-1.72(3) & -1.49(4)   & 0.44(1) & 3.81(1) & 1.74(1) & 0.78(8) & 25.53(6) & 6.05(1) \\
    V3   &-1.67(3) & -1.63(3)   & 0.51(1) & 3.81(1) & 1.71(1) & 0.73(7) & 23.60(3) & 5.86(1) \\
    V7   &-1.74(3) & -1.72(3)   & 0.59(1) & 3.81(1) & 1.68(1) & 0.72(7) & 24.34(3) & 5.70(1) \\
    V13   &-1.83(4) & -1.84(5)   & 0.48(1) & 3.80(1) & 1.73(1) & 0.74(8) & 23.70(5) & 6.13(1) \\
    \hline
   Weighted Mean &-1.67 & -1.62  & 0.52 & 3.80 & 1.71& 0.72 & 23.85 & 5.89\\
        $\sigma$ & $\pm 0.11$  &$\pm 0.15$ & $\pm 0.05$ & $\pm 0.01$ & $\pm 0.01$  & $\pm 0.04$ &  $\pm 0.84$ & $\pm 0.14$ \\
    \hline
    \multicolumn{8}{c}{RRc}\\
    \hline
   V4 &-1.37(22)  & -1.26(22)   & 0.56(4) & 3.86(1) & 1.68(1)  & 0.43(1)  & 24.00(54) & 4.37(9) \\
   
V5 &-2.07(8)  & -2.20(12)   & 0.41(4) & 3.85(1) & 1.74(1)  & 0.54(1)  & 24.46(21) & 4.93(9) \\

    V6 &-1.79(4)  & -1.78(5)   & 0.54(2) & 3.86(1) & 1.68(1)  & 0.53(1)  & 24.46(20) & 4.43(4) \\
    
   V8 &-1.79(9)  & -1.78(12)   & 0.55(6) & 3.86(1) & 1.68(2)  & 0.55(4)  & 24.59(63) & 4.40(11) \\
      
    V17 &-1.59(3)  & -1.51(3)   & 0.46(1) & 3.86(1) & 1.72(1)  & 0.47(1)  & 24.03(12) & 4.66(2) \\
    V18 &-1.67(17) & -1.62(20)   & 0.59(1) & 3.86(1) & 1.67(1)  & 0.52(3)  & 22.02(9) & 4.28(2) \\
     V19 &-2.00(14)  & -2.10(21)   & 0.59(3) & 3.86(1) & 1.67(1)  & 0.64(4)  & 22.22(16) & 4.39(3) \\   
    \hline
    Weighted Mean & -1.69 &	-1.62 & 0.54 & 3.86 &  1.68 & 0.49& 22.90  & 4.42\\
    
    $\sigma$ & $\pm 0.22$  &$\pm 0.30$ & $\pm 0.06$ & $\pm 0.01$ & $\pm 0.01$  & $\pm 0.01$ &  $\pm 1.01$ & $\pm 0.25$ \\
   \hline
\end{tabular}
\label{tab:parfisRR}
\end{table*}

The average [Fe/H] and $M_V$ (hence distance), should be representative of the parent cluster. The iron abundance is given in two scales, [Fe/H]$_{\rm ZW}$, in the scale of \citet{Zinn1984}
which can be transformed into the spectroscopic scale of \citet{Carretta2009}, via the equation; [Fe/H]$_{\rm UVES}$= $-0.413$ + 0.130~[Fe/H]$_{\rm
ZW} - 0.356$~[Fe/H]$_{\rm ZW}^2$.

Given the Galactic position of the cluster, it is subject to very low dust extinction. We have adopted $E(B-V)= 0.05$ \citep{Harris1996,Schlafly2011,Schlegel1998}.

We found averages [Fe/H]$_{\rm ZW}= -1.67 \pm 0.11$, [Fe/H]$_{\rm UV}= -1.62 \pm 0.15$ and $D=23.9 \pm 0.8$ kpc from the RRab light curves
and [Fe/H]$_{\rm ZW}= -1.69 \pm 0.22$, [Fe/H]$_{\rm UV}= -1.62 ~\pm~0.30$ and $D=22.9 \pm 1.0$ kpc from the RRc data. The results for the 5 RRab and 7 RRc stars come from independent calibrations, obtained with different sets of calibrators, hence, results are also independent and agree very well.
These values can be compared with the value quoted [Fe/H]$_{Spec}=-1.93 \pm 0.09$ in the spectroscopic scale \citep{Carretta2009} and with the canonical value [Fe/H]=$-1.88 $ listed by \citet{Harris1996}, and the mean distance 
$D=25.959 \pm 0.62$ kpc estimated by \citet{Baumgardt2021}. It is worth to note that the value of the metallicity  reported by APOGEE DR17 \citep{Schiavon2024} is [Fe/H]= $-1.72$, obtained for 2 cluster stars. Despite this, our results show a slightly metal richer and closer cluster, however we stress that the Fourier average results in Table \ref{tab:parfisRR} were limited to clear cluster members that are well placed on the Horizontal Branch as shall be shown in the following section.

\section{The Colour-Magnitude Diagram}

The $V -(V-R)$ Colour-Magnitude diagram of NGC 5634 is shown in Fig \ref{CMD}. Red points are stars found as likely cluster members, while gray dots represent the field star. Blue and green circles are used for RRab and RRc stars, respectively. Hence, the cluster member variables have a red dot in the center. The position of V10 is spurious; the star is blended, and we were unable to resolve it. The stars V7, V9, V12, V15 and V20, which according to their proper motions are likely cluster members, appear way above the HB, however, as discussed in Section \ref{par:Bailey} for V12 and V20, $Gaia$ sources within 2 arc seconds were also found for V7, V9 and V15. After correcting for the presence of unresolved neighbours for these five stars, they become closer to the HB (crosses in Fig. \ref{CMD}). Their corrected $<V>$ magnitudes were used in the calculation of their distance in the Fourier analysis in Section \ref{sec:Four}.

The Fourier physical parameters listed in Table \ref{tab:parfisRR} were calculated only for properly resolved cluster member RR Lyrae stars.

\section{Summary of results}

We were able to recover good light curves of some of the RR Lyrae in the field of NGC 5634, a few of them being well inside the central region of the cluster. A dedicated membership analysis employing the method of \citet{Bustos2019} and the independent analysis of \citet{Vasiliev2021} \citep[see also][]{Prudil2024} find that V11 and V16 do not pertain to the cluster. For V20 the two approaches find the star to be a member and a field star respectively. Membership methods are always subject to statistical considerations and may disagree on individual star, particularly in crowded regions, as it is the case of V20. This RRab star is badly blended with a brighter star which causes the star to appear brighter and with a smaller amplitude. Correcting by the presence of the contaminant star, V20 falls near to the HB and to the OoII lucus on the Bailey's diagram. Thus we have classified the star as a likely cluster member.

The periods and equatorial coordinates of the variable stars have been refined and are reported in Table \ref{tab:datosgenerales1}. The coordinates listed in the table correspond to the place where the detailed blinking comparison of the differential images marked the variations; they may differ a bit from the coordinates listed in the CVSGC. Accurate identifications are given in the charts of Fig. \ref{ID_CHART}.

The mean cluster iron abundance and distance found from independent calibrations of the Fourier decomposition parameters for the RRab and RRc stars light curves are [Fe/H]$_{\rm ZW}= -1.68 \pm 0.18$, [Fe/H]$_{\rm UV}= -1.62 \pm 0.26$ and $D=23.4 \pm 0.9$ kpc. 

\vskip 1.0cm

\section*{Acknowledgments}
The comments and suggestions from an anonymous referee are deeply appreciated as they have contributed to improve the manuscript. AAF is grateful to the Indian Institute of Astrophysics, for warm hospitality during the observations at the IAO from the Hosakote (CREST) station.  The authors thank the SPM and IAO Telescope Allocation Committees for the granted observing time, and to the technical personnel in both observatories for making possible the data acquisition.
The present project has been partially supported by DGAPA-UNAM through project  IN103024. 
The IAO and CREST facilities are operated by the Indian Institute of Astrophysics, Bangalore.

This work has made use of data from the European Space Agency (ESA) mission
{\it Gaia} (\url{https://www.cosmos.esa.int/gaia}), processed by the {\it Gaia}
Data Processing and Analysis Consortium (DPAC,
\url{https://www.cosmos.esa.int/web/gaia/dpac/consortium}). Funding for the DPAC
has been provided by national institutions, in particular the institutions
participating in the {\it Gaia} Multilateral Agreement.

\section*{DATA AVAILABILITY}
The data employed in this work shall be available in electronic
form in the Centre de Donnés astronomiques de Strasbourg data
base (CDS), and can also be shared on request to the corresponding
author.

\bibliographystyle{mnras}
\bibliography{NGC5634}

\bsp	
\label{lastpage}
\end{document}